\documentclass[twocolumn,preprintnumbers,superscriptaddress,amsmath,amssymb,prb]{revtex4-1}

\usepackage{graphicx}
\usepackage{dcolumn}
\usepackage{subfigure}
\usepackage{bm}
\usepackage[utf8]{inputenc}  
\usepackage{comment}
\usepackage{units}
\usepackage[unicode=true, bookmarks=false, breaklinks=false, pdfborder={0 0 1},colorlinks=false]{hyperref}
\usepackage{breakurl}
\usepackage{url}
\usepackage{natbib}
\usepackage{multirow}
\usepackage{float}
\usepackage{verbatim}
\usepackage{booktabs}

\usepackage[cmyk,dvipsnames]{xcolor}
\definecolor{pacificb}{HTML}{1CA9C9}

\begin{document}

\title{Ultrafast demagnetization dynamics of 4\textit{f} antiferromagnets}

\newcommand{\KTH}{Department of Applied Physics, School of Engineering Sciences, KTH Royal Institute of Technology, 
AlbaNova University Center, SE-10691 Stockholm, Sweden}
\newcommand{\SeRC}{SeRC (Swedish e-Science Research Center), KTH Royal Institute of Technology, SE-10044 Stockholm, Sweden}
\newcommand{\Uppsala}{Department of Physics and Astronomy, Uppsala University, Box 516, SE-75120 Uppsala, Sweden}
\newcommand{\WISEUppsala}{Wallenberg Initiative Materials Science for Sustainability, Uppsala University, 75121 Uppsala, Sweden}
\newcommand{\Skovde}{Department of Engineering Sciences, University of Skövde, SE-541 28 Skövde, Sweden}

\newcommand{\Orebro}{School of Science and Technology, \"Orebro University, SE-701 82, \"Orebro, Sweden}
\newcommand{\Stockholm}{Department of Materials and Environmental Chemistry, Stockholm University, SE-10691 Stockholm, Sweden}
\newcommand{\UppsalaChem}{Department of Chemistry - Ångström Laboratory, Uppsala University, Box 538, Uppsala, SE-751 21, Sweden}
\newcommand{\Berlin}{Department  of  Physical  Chemistry,  Fritz  Haber  Institute  of  the  Max  Planck  Society,  Faradayweg  4-6,  14195 Berlin, Germany}
\newcommand{\Linz}{Institute for Theoretical Physics, Johannes Kepler University, Altenberger Strasse 69, 4040 Linz, Austria}
\newcommand{\Halle}{Max-Planck-Institut für Mikrostrukturphysik, Weinberg 2, 06120 Halle (Saale), Germany}

\author{Maryna Pankratova}
\email[Corresponding author:\ ]{maryna.pankratova@his.se}
\affiliation{\Skovde}
\affiliation{\Uppsala}
\affiliation{\WISEUppsala}

\author{Vladislav Borisov}
    \affiliation{\Uppsala}
   
\author{Danny Thonig}
    \affiliation{\Orebro}
    \affiliation{\Uppsala}

\author{Rohit Pathak}
  \affiliation{\Uppsala}

\author{Yoav William Windsor}
    \affiliation{\Berlin}    

\author{Laurenz Rettig}
    \affiliation{\Berlin}

\author{Olle Eriksson}
    \affiliation{\Uppsala}
\affiliation{\WISEUppsala}

\author{Arthur Ernst}
    \affiliation{\Linz}
    \affiliation{\Halle} 

\author{Anders Bergman}
\affiliation{\Uppsala}

\date{\today}

\begin{abstract}
We study the ultrafast demagnetization dynamics of  \textit{Ln}Rh$_2$Si$_2$ (\textit{Ln} $=$ Pr, Nd, Sm, Gd, Tb, Dy, Ho)
antiferromagnets after excitation by a laser pulse, using a combination of density functional theory and atomistic spin and spin-lattice dynamics simulations. In the first step, we calculate the Heisenberg interactions using the magnetic force theorem and compare two approaches, where the $4f$ states of the rare earths are treated as frozen core states or as valence states with added correlation corrections. We find marked quantitative differences in terms of predicted Curie temperature for most of the systems, especially for those with large orbital moment of the rare earth cations. This can be attributed to the importance of indirect interactions of the $4f$ states through the Si states, which depends on the binding energy of the $4f$ states and coexists with RKKY-type interactions mediated by the conduction states. However, qualitatively both approaches agree in terms of the predicted antiferromagnetic ordering at low temperature, which is in line with previous experiments. In the second step, the atomistic dynamics simulations are used in combination with a heat-conserving two-temperature model, that allows for the calculation of spin and electronic temperatures during the magnetization dynamics simulations. Our simulations demonstrate that despite quite different demagnetization times, magnetization dynamics of all studied \textit{Ln}Rh$_2$Si$_2$ antiferromagnets exhibit similar two-step behavior, in particular,  the first fast drop followed by slower demagnetization. 
In addition, we observe that the demagnetization amplitude depends linearly on laser fluence, for low fluences, something which also is in agreement with experimental observations. We also investigate the impact of lattice dynamics on ultrafast demagnetisation using coupled atomistic spin-lattice dynamics simulations and heat-conserving three temperature model, which confirm linear dependence of magnetisation on laser fluence. The microscopic mechanisms behind these behaviors are here investigated in detail. 
\end{abstract}

\maketitle

\section{Introduction}

The field of ultrafast magnetization dynamics was launched in 1996 by the work of Beaurepaire and coauthors \onlinecite{beaurepaire1996ultrafast}, who studied the impact of the femtosecond laser pulse on the demagnetization of Ni films and demonstrated for the first time ultrafast demagnetization on subpicosecond timescales. Ultrafast demagnetization can be very useful for prospective spintronic devices since it could allow switching of magnetization on picosecond timescales, addressing the demand for a growing data storage speed. While, ultrafast demagnetization was first observed in nickel \onlinecite{beaurepaire1996ultrafast,PhysRevB.106.174407}, numerous studies have followed, including  ferromagnetic (FM) \onlinecite{PhysRevResearch.3.023032,PhysRevB.107.L100407,doi:10.1126/sciadv.abb1601,PhysRevB.91.014425, pankratova2024coupled, PhysRevB.107.174431}, and ferrimagnetic materials \onlinecite{gupta2024tuning}.  However, for spintronic applications, antiferromagnetic (AFM) materials form perhaps an even more promising class of magnets \onlinecite{windsor2022exchange,PhysRevResearch.6.043019,lee2022robust}, due to properties such as robustness against external magnetic perturbations. In addition, an experimental study comparing FM and AFM ultrafast demagnetization for Dy report that quenching of magnetization is faster and more efficient for the AFM phase than for the FM one \onlinecite{PhysRevLett.119.197202}. Moreover, while comparing FM and AFM order between FeGd sublattices, it was theoretically predicted that AFM coupling of the sublattices accelerates the demagnetization of both sublattices \onlinecite{PhysRevLett.108.057202}.

Recently, the ultrafast demagnetization of a series of lanthanide-based $4f$ antiferromagnets was systematically experimentally studied in Ref.~\onlinecite{windsor2022exchange}. The considered materials were prepared specifically to possess similar lattice and magnetic structure, allowing to focus solely on the impact of $4f$ occupation. It was demonstrated that the timescales of the demagnetization dynamics can differ up to two orders of magnitudes, however, the angular momentum transfer rates scale linearly with the de Gennes factor $G = (g_J – 1)^2J(J + 1)$, where $g_J$ and $J$ are the Landé factor and the total $4f$ angular momentum quantum number, respectively.  In addition, density functional theory (DFT) calculations were performed in Ref.~\onlinecite{windsor2022exchange} to obtain the corresponding exchange interaction values. 
Details on the correct description and methodology for the localized $4f$ electrons was not provided in Ref.~\onlinecite{windsor2022exchange} and will be discussed here.

In Ref.~\onlinecite{windsor2022exchange}, it was found that coupling between the nearest antiferromagnetically aligned spins also scales linearly with the de Gennes factor. Moreover, it was shown that the demagnetization amplitude increases linearly with laser fluence for all studied systems, which agrees with previous experimental observations for other materials (see Refs. \onlinecite{lee2024controlling,pankratova2024coupled} and references therein). It should be noted that magnetization dynamics studies were not performed in Ref.~\onlinecite{windsor2022exchange} to confirm that the simulated demagnetization amplitude based on such interactions depends linearly on laser fluence, and filling in this theoretical gap is in the focus of the present work.

Theoretically, ultrafast magnetization dynamics is often interpreted using two- or three-temperature models \onlinecite{beaurepaire1996ultrafast,lee2022robust, PhysRevB.106.174407}. These models usually include two or three coupled reservoirs, i.e. spin-, electron-, and lattice. In the three-temperature model (3TM) proposed by Beaurepaire \onlinecite{beaurepaire1996ultrafast}, there are several coupling constants that are responsible for the transfer of heat between the subsystems: electron-phonon G$_{ep}$, electron-spin G$_{es}$, and spin-lattice G$_{sl}$. These constants for 3d ferromagnets, such as iron, cobalt, and nickel are estimated in various works with up to one order of magnitude difference \onlinecite{PhysRevResearch.3.023032,PhysRevB.106.174407,pankratova2024coupled}, which impede the theoretical interpretations of experimental observations. In addition, the 3TM proposed by Beaurepaire fails to describe demagnetization on subpicosecond timescales. Recently, a heat-conserving three-temperature model (HC3TM) was proposed \onlinecite{PhysRevB.106.174407}, where Gilbert and lattice dampings govern a coupling between the electron, lattice, and spin degrees of freedom. These values can be calculated from ab-initio, and can thus facilitate a comparison between experiment and calculations more accurately and is in addition less dependent on empirical parameters. Moreover, the HC3TM results in a qualitatively correct description of the demagnetization dynamics on subpicosecond timescales for all 3d ferromagnets \onlinecite{PhysRevB.106.174407,pankratova2024coupled}. Building on these successes, the HC3TM is a promising candidate for a more realistic description of demagnetization times on sub-picosecond timescales also in the \textit{Ln}Rh$_2$Si$_2$ series, where the quite significant differences in demagnetization times, especially the fast demagnetization of SmRh$_2$Si$_2$, are making the 3TM of Ref. \onlinecite{beaurepaire1996ultrafast} questionable, especially facing the difficulty of identifying several coupling constants needed in this model; G$_{ep}$, G$_{es}$, G$_{sl}$. 

In this work, we report DFT calculations of exchange interaction parameters for \textit{Ln}Rh$_2$Si$_2$ and atomistic spin dynamics simulations of ultrafast demagnetization dynamics combined with the heat-conserving two and three-temperature models.
The paper is organized as follows. First, we present the results of density functional theory calculations. Then, the calculated parameters are used in atomistic spin and spin-lattice dynamics simulations to study ultrafast demagnetization dynamics and the impact of laser fluence and lattice dynamics on magnetization dynamics of lanthanide-based antiferromagnets. 

\section{Methods}

\subsection{First principles density functional theory}
In the first step, we calculate the electronic properties of the rare earth compounds LnRh$_2$Si$_2$ based on the measured crystal structure, using two different approaches: \textit{i)} all-electron full-potential fully relativistic electronic structure software RSPt [\onlinecite{Wills1987},\onlinecite{Wills2010}] and \textit{ii)} a full-relativistic multiple scattering Green's function method theory formulated in the Korringa-Kohn-Rostoker approach and implemented in the code \textit{HUTSEPOT}[\onlinecite{Geilhufe2015},\onlinecite{Hoffmann2020}].

In \textit{i)}, the electronic structure is described on the level of density functional theory within the generalized-gradient approximation in PBE parametrization. Smearing of $\unit[1]{mRy}$ is used for electronic occupations, and a $(40\times 40\times 16)$ $k$-mesh is used for sampling the Brillouin zone. An AFM structure is obtained in our calculations, where each \textit{Ln} layer is ferromagnetic but neighboring layers are AFM-coupled, which is close to the measured magnetic ground states of the studied rare-earth compounds. We model these systems by following the standard model for rare earth compounds, i.e. that the 4$f$ states of the rare earth cations are considered as core states with zero dispersion and a fixed magnetic moment corresponding to the Hund's rules in the atomic picture. These 4$f$ core states affect the other electronic states, e.g.~spin-polarize the $s$, $p$ and $d$ states of the same cations, which is relevant for determining the magnetic interactions (see further below). Importantly, the muffin-tin radius of the \textit{Ln} cations is set as large as possible at $\unit[3.3]{a.u.}$, which affects the calculated muffin-tin magnetic moments and their interactions due to the spatial extent of the $s$, $p$ and $d$ states mentioned before.

In \textit{ii)}, strongly localized $4f$ electrons of the \textit{Ln}Rh$_2$Si$_2$ compounds were treated as valence electrons and within the GGA + U approach \onlinecite{Anisimov1991}. The corresponding effective Hubbard parameter $U_\mathrm{eff}=U-J=6$ eV was chosen in such a way as to guarantee a good agreement of calculated and experimental N\'eel temperature.

Both methods, \textit{i)} and \textit{ii)}, provide similar electronic and magnetic structures of the here studied compounds, as discussed in the following.

\subsection{LKAG formula}

We calculated the Heisenberg magnetic interactions between the rare earth spins using the Liechtenstein-Katsnelson-Antropov-Gubanov (LKAG) approach [\onlinecite{LKAG1987}] (for a recent review, see \onlinecite{Szilva2023}) using the Full-Potential Linear Muffin-tin Orbital methods implemented in the RSPt software [\onlinecite{Wills1987},\onlinecite{Wills2010}] as well as the fully relativistic Korringa Kohn Rostoker Green function method implemented  in the code \textit{HUTSEPOT} ~[\onlinecite{Geilhufe2015},\onlinecite{Hoffmann2020}]. The LKAG approach has an advantage compared to other methods in the literature, such as the total energy mapping, since it allows to calculate interactions for distances of up to several lattice parameters without using excessively large supercells, which reduces significantly the computational time. It should be noted that, since we put the $4f$ states of the rare earth in the core in the RSPt calculations, the Heisenberg exchange is calculated including the $s$, $p$ and $d$ states of the same cations which are polarized due to large $4f$ moments. In the \textit{HUTSEPOT} calculations, the $4f$ states are treated as valence states using either GGA+$U$ or self-interaction corrections (SIC), so the magnetic interactions between the rare-earth cations are calculated taking into account not only $s$, $p$ and $d$ but also $f$ states.

The magnetic energy of the system is described in this case by the classical Hamiltonian:
\begin{equation}
    H = -\frac12 \sum\limits_{j\neq i} J_{ij} \boldsymbol{e}_i \cdot \boldsymbol{e}_j.
    \label{e:Heisenberg_model}
\end{equation}
Here, the unit vectors $\boldsymbol{e}_i=\boldsymbol{m}_i/m_i$ describes the direction of magnetic moment on each rare earth cation at site $i$. Due to crystal symmetry the Dzyaloshinskii-Moriya interactions in the studied systems are zero and, for that reason, are not considered in the spin Hamiltonian (\ref{e:Heisenberg_model}). Also, we disregard relativistic corrections to the electronic structure when calculating the magnetic interactions in \textit{Ln}Rh$_2$Si$_2$, but we discuss the effect of such corrections on the density of states in Sec.~IV A. Importantly, we find that the calculated exchange parameters $J_{ij}$ are quite sensitive to the calculation scheme. Both $J_{ij}$ parameter sets (from RSPt and \textit{HUTSEPOT}) provide the correct magnetic ground state, but, since \textit{HUTSEPOT} values of the $J_{ij}$ parameters provide better estimates of the magnetic ordering temperature, we use those values for the spin dynamics simulations in the two- and three-temperature model (see next section).

\section{Atomistic magnetization dynamics and temperature model}

In atomistic spin dynamics simulations \onlinecite{UppASD}, the time evolution of atomic spins is governed by the Landau-Lifshitz-Gilbert equation \onlinecite{Landau1935,Gilbert2004}, that can be expressed in Landau-Lifshitz form:
\begin{align}\label{eq1}
\frac{d \boldsymbol{m}_i}{d t}
& = - \frac{\gamma}{(1+\alpha ^2)}\boldsymbol{m}_i\times(\boldsymbol{B}_i+\boldsymbol{B}_i^{\mathrm{fl}})\\
& \quad - \frac{\gamma}{(1+\alpha ^2)}\frac{\alpha}{m_i}\boldsymbol{m}_i\times(\boldsymbol{m}_i \times [\boldsymbol{B}_i+\boldsymbol{B}_i^{\mathrm{fl}}]),\notag
\end{align} 
where $m_i$ is the saturation magnetization and $\gamma$ represents the gyromagnetic ratio, and $\alpha$ is the Gilbert damping. An effective exchange field $\boldsymbol{B}_i = - \partial H/\partial \boldsymbol{m}_i$ can be obtained from the spin Hamiltonian \eqref{e:Heisenberg_model}, where the exchange parameters $J_{ij}$ from the \textit{HUTSEPOT} code are used. In our simulations, temperature is included by means of Langevin dynamics where we introduce a stochastic field $\boldsymbol{B}_i^\mathrm{fl}$ as white noise with properties $\langle B_{i,\mu}^\mathrm{fl}(t) B_{j,\nu}^\mathrm{fl}(t') \rangle=2D^M_i \delta_{ij}\delta_{\mu\nu}\delta(t-t')$ with $\mu,\nu=x,y,z$. In particular, in these calculations we employ $D^M_i= \alpha k_B T_e/(1+\alpha^2)\gamma m_i$, where $T_e$ and $k_B$ are electronic temperature and Boltzmann constant respectively (see Ref.~\onlinecite{Eriksson2017}).
The formalism above is available in the UppASD\- \onlinecite{Eriksson2017} code which was used for all simulations in this work. In our simulations of ultrafast demagnetisation dynamics, we employ a simulation cells with a $40 \times 40 \times 40$ repetition of the unit cell, using periodic boundary conditions.  In addition, $N_t = 10^6$ time steps of $dt = \unit[10^{-16}]{s}$ were used.

\begin{figure}[h]
\includegraphics[width=0.37\textwidth]{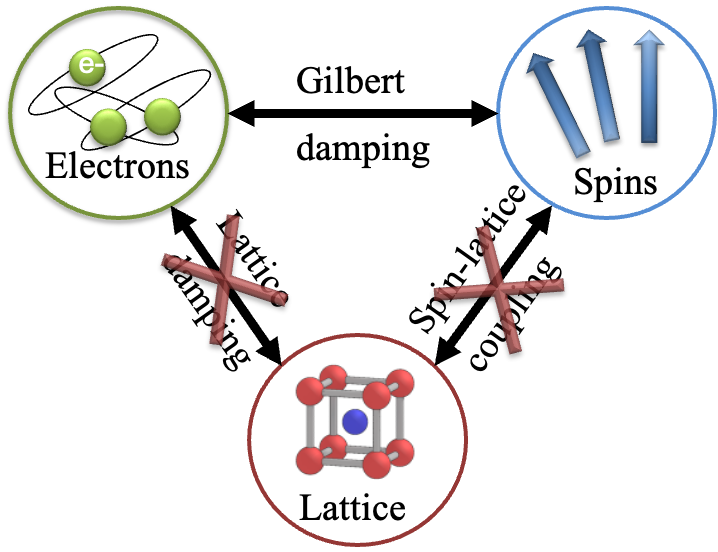}
\caption{Schematic of the HC2TM model. \label{fig:HC2TM}}
\end{figure}
In this work, we reduce the HC3TM to calculate electron and spin temperatures in ultrafast magnetization dynamics simulations. We choose to exclude lattice temperature from consideration since it is stated in Ref.\onlinecite{windsor2022exchange} that materials under investigation here have very similar $4f$ spin arrangement, crystal structure, and chemical bonding characteristics with only $2.5 \%$ difference in lattice constants. It is to be noted that the only difference between them is the occupation of the localized $4f$ shell of the \textit{Ln} ions, and hence the magnetic moment. Therefore, it is reasonable to study whether the difference between the studied systems is only due to the exchange interactions, and whether it is sufficient to reproduce experimental findings without taking the lattice part into account. We attempt to answer this question in the present work.

In this work we primarily use the HC2TM, with only spin and electron temperatures considered, but we make a detailed comparison between HC2TM and HC3TM for one of the compounds considered here.  
In the HC2TM, we modify the expression for the time-dependent electronic temperature dynamics in the HC3TM, proposed in Ref.~\onlinecite{PhysRevB.106.174407}, to study a case of a fixed lattice temperature, and obtain:

\begin{equation}
    \Delta T_e (t) = - \frac{C_s(T_s)}{C_e(T_e)}\Delta T_s(t) + \frac{W(t)}{C_e(T_e)}-\frac{G_\mathrm{cool}(T_e-T_\mathrm{final})}{C_e(T_e)},
    \label{eq5}
\end{equation}
Here, the spin temperature is defined by the expression $T_s=\frac{m\langle\sum_{i}\left|\boldsymbol{e}_{i}\times\boldsymbol{B}_i\right|^2\rangle}{2k_B\langle\sum_{i}\boldsymbol{e}_{i}\cdot\boldsymbol{B}_i\rangle}$, where $\boldsymbol{e}_{i}$ is the normalized local spin moment \onlinecite{PhysRevE.82.031111}. The details of calculating spin and electronic temperatures in HC3TM can be found in Ref.~\onlinecite{PhysRevB.106.174407}. Also, $C_e(t)$ and $C_s(t)$ are electron and spin heat capacities correspondingly. The laser impact is captured by the term $W(t)$ which is modeled as a Gaussian. $G_{cool}$ corresponds to heat dissipation from the laser effected spot to the whole sample. This term is relevant in this study because of the rather big time-scales of demagnetization (up to 400 ps for holmium). Moreover, this term can serve as an additional cooling channel partly substituting a full treatment of the lattice subsystem. Here, we choose the value of $G_\mathrm{cool}$ so the remagnetization is close to the experimentally observed one. The value of $G_\mathrm{cool}=2\times 10^{16} $ J/s is the same for all studied here materials.

\section{Results}
\subsection{Electronic structure}

\begin{figure*}
    \includegraphics[width=0.99\textwidth]{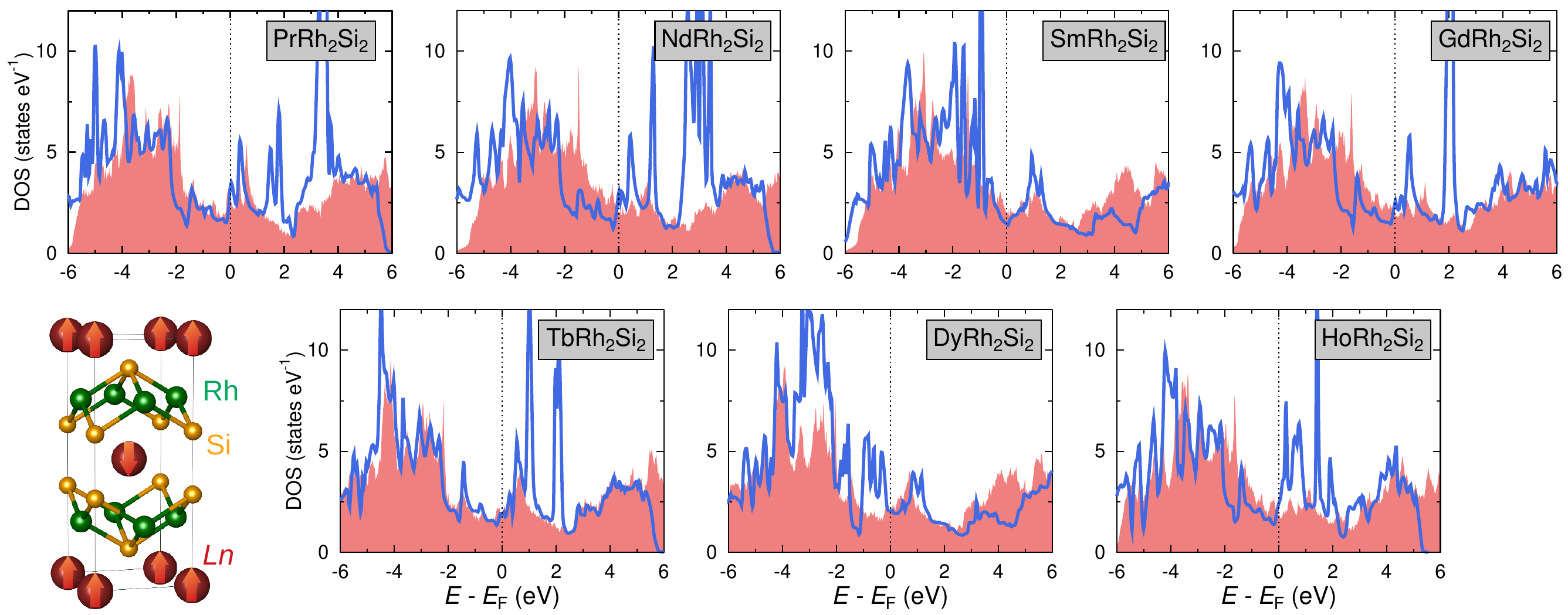}
    \caption{Density of states calculated for the antiferromagnetic state of different rare-earth compounds \textit{Ln}Rh$_2$Si$_2$ (\textit{Ln} = Pr, Nd, Sm, Gd, Tb, Dy, Ho) using scalar relativistic $4f$ in the core (red filling) and full-relativistic GGA+$U$ method (blue lines).}
    \label{fig:DOS_LnRh2Si2}
\end{figure*}

The electronic structure of \textit{Ln}Rh$_2$Si$_2$ was examined by both scalar- and full-relativistic approaches. Although $4f$ elements were treated as core electrons in our scalar-relativistic RSPt calculations, the densities of states (DOS) at the Fermi level  are very similar in both approaches (see Fig.~\ref{fig:DOS_LnRh2Si2}). This is crucial for the RKKY interaction, which is a dominant interaction between the localized 4$f$ magnetic moments in these systems. However, the DOSs in the range of the valence bands are different in both approaches in the case where the total angular moment of the $4f$ atom is larger than zero. The most prominent deviations occur for Dy and Ho cases, since the spin-orbital interaction and other relativistic effects are significant in these systems.
However, since 4$f$ states remain to be sufficiently localized, their positions do not affect the DOS at the Fermi level, which is mainly of 5$d$ character. Nevertheless,  the position of $4f$ states can be important for other types of interaction between the magnetic moments, for example the indirect double exchange interaction via Si $sp$ states. Thus, although treating $4f$ states as core might predict correctly the ground state, since the RKKY interaction is dominant in these systems, the critical temperature might be estimated wrongly by this approach. 

The obtained magnetic moments are presented in Table I. In the scalar-relativistic calculations, the spin moment of the $4f$ states is fixed at the value obtained from the Russell-Saunders rules, while the contribution from the $s$-, $p$- and $d$-states is calculated self-consistently using DFT. This $spd$-contribution is similar both in scalar- and full-relativistic calculations for some of the compounds (e.g.~Tb-, Ho- and Pr-based) but can be visibly different for others (e.g.~Gd and Dy cases), as one can see from the middle column of Table~I. Relativistic calculations with $4f$ electrons in the valence parts provide also orbital moments, which are substantial for all elements except for Gd. Nevertheless, the magnitude of the DOS at the Fermi level is almost the same in both approaches. This is crucial for the RKKY interaction, which is dominant in in this family of  \textit{Ln}Rh$_2$Si$_2$ compounds. Therefore, both scalar- and full-relativistic methods should deliver a similar behavior for Heisenberg exchange constants, which are discussed in the next subsection. We note that the deviations of the $spd$-moments can partly explain the quantitative differences in the Heisenberg interactions that we obtain using the two calculation methods.

\setlength{\tabcolsep}{5pt}
\renewcommand{\arraystretch}{1.5}
\begin{table}[h]
 \caption{Calculated magnetic $m^s$ and orbital moments $m^l$ of the rare-earth \textit{Ln} cations in \textit{Ln}Rh$_2$Si$_2$ compounds. The left and right parts of the middle columns show the results where the 4\textit{f} states of \textit{Ln} cations are set as the core states (``core'') and from full-relativistic self-interaction corrected treatment (``SIC'', respectively. 
 The de-Gennes factor $G = (g - 1)^2 J (J+1)$ together with the interlayer Heisenberg exchange $J_3$ calculated from HUTSEPOT (second last column) and RSPt (last column) are indicated as well.} 
\vspace{5pt}
  \centering
  \begin{tabular}{c|cc|cc|p{0.7cm}|p{0.7cm}p{0.7cm}p{0.7cm}}
    \hline
    \textit{Ln} & \multicolumn{2}{c|}{$m^s_{4f}\:(\mu_\mathrm{B})$} & \multicolumn{2}{c|}{$m^s_{spd}\:(\mu_\mathrm{B})$} & $m^l_f$ $(\mu_\mathrm{B})$ & $G$ & $J_3$ ($\mathrm{\mu}$Ry)& $J_3$ ($\mathrm{\mu}$Ry) \\
    \hline
       & core & SIC & core & SIC & SIC & & & \\ \hline
    Pr & 1.60 & 0.63 &0.07 & 0.06 & 3.25 & 0.80 &  -5.6 & -5.5 \\
    Nd & 2.45 & 2.54 &0.08 & 0.14 & 5.81 & 1.84 &  -4.1 & -5.3 \\
    Sm & 3.43 & 2.88 &0.10 & 0.08 & 1.52 & 4.11 &  -6.0 & -8.1 \\
    Gd & 7.00 & 7.03 &0.18 & 0.30 & 0.00 & 15.8 & -15.7 & -23.7 \\
    Tb & 6.00 & 5.58 &0.19 & 0.22 & 2.27 & 10.5 & -10.8 & -31.4 \\
    Dy & 5.00 & 3.63 &0.15 & 0.09 & 6.02 & 7.08 & -6.8 & -12.9 \\
    Ho & 4.00 & 3.24 &0.11 & 0.12 & 5.42 & 4.50 &  -6.7 & -5.0 \\
    \hline
  \end{tabular}
  \label{t:magnetic_moments}
\end{table}

\subsection{Exchange coupling constants}

The Heisenberg exchange parameters calculated using the RSPt code for \textit{Ln}Rh$_2$Si$_2$ show certain characteristic trends. In particular, the interlayer exchange $J_3$ which stabilizes the AFM-coupled rare-earth layers scales mostly linearly with the de-Gennes factor. The largest deviations from linearity are observed for \textit{Ln} = Pr and Tb. This linear trend is also discussed in previous work \onlinecite{windsor2022exchange} where the magnetic interactions were calculated using both LMTO and KKR methods, and the latter show a similar but more linear trend. Overall, these results suggest a physical picture where the magnetic exchange between the rare-earth moments is mediated by conduction electrons, leading to RKKY-like character of interactions and linear scaling with the de-Gennes factor, already discussed in depth in Ref. \onlinecite{windsor2022exchange}.\\

\subsection{Magnetic ground state}

\begin{figure}[htb]
\includegraphics[width=0.45\textwidth]{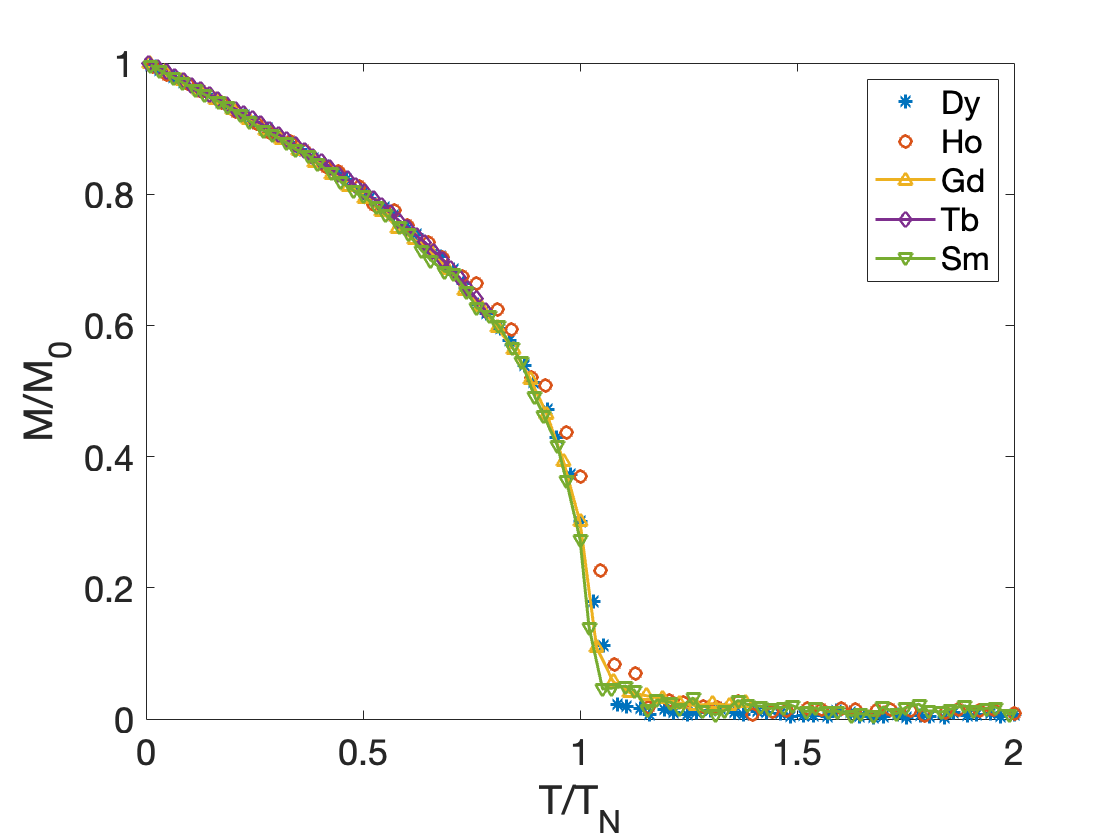}
\caption{Normalized magnetization M/M$_0$ as a function of normalized temperature for \textit{Ln}Rh$_2$Si$_2$ materials with different \textit{Ln} elements specified in the figure legend. \label{fig:M_temp} }
\end{figure}

\begin{figure}[htb]
\includegraphics[width=0.45\textwidth]{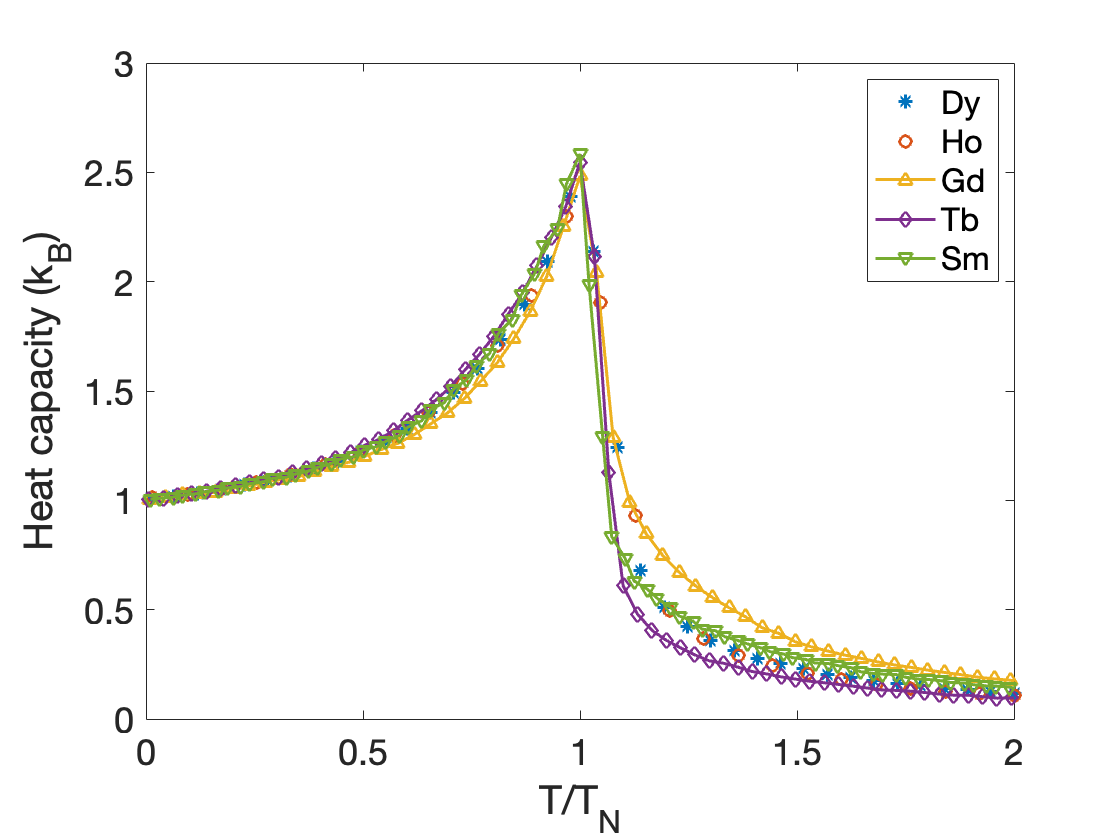}
\caption{Temperature-dependent spin heat capacities for \textit{Ln}Rh$_2$Si$_2$ materials with different \textit{Ln} cations specified in the figure legend. The temperatures are normalized to the corresponding N\'eel temperatures T$_N$.}\label{fig:C_temp}
\end{figure}

Following the experimental studies  \onlinecite{windsor2022exchange}, we start the analysis of the magnetic properties of the system by performing calculations of the temperature dependence of the magnetization, using the calculated exchange constants discussed above. The results of these calculations are presented in Fig.~\ref{fig:M_temp}, where the average magnetization curves for one of the
AFM sublattices are shown. It can be seen from the figure, that, despite very different magnetic moments and N\'eel temperatures, the normalized magnetization as a function of normalized temperature behaves similarly for all studied materials, in agreement with experimental observations \onlinecite{windsor2022exchange}. In addition, in Fig.~\ref{fig:C_temp} we also present the heat capacities of the spin system, which will be further used for calculations of ultrafast demagnetization dynamics in HC2TM, which was not done for these systems in the literature so far to the best of our knowledge. In general, the importance of taking into account temperature-dependent heat capacities in HC3TM was discussed in Ref.~\onlinecite{PhysRevB.106.174407}. In calculations of spin heat capacities, we employ the widely used classical approach based on Boltzmann statistics (please see Refs. \onlinecite{PhysRevMaterials.2.013802,PhysRevB.106.174407} and references therein).

\subsection{Ultrafast demagnetization}

\begin{figure}[h]
\includegraphics[width=0.45\textwidth]{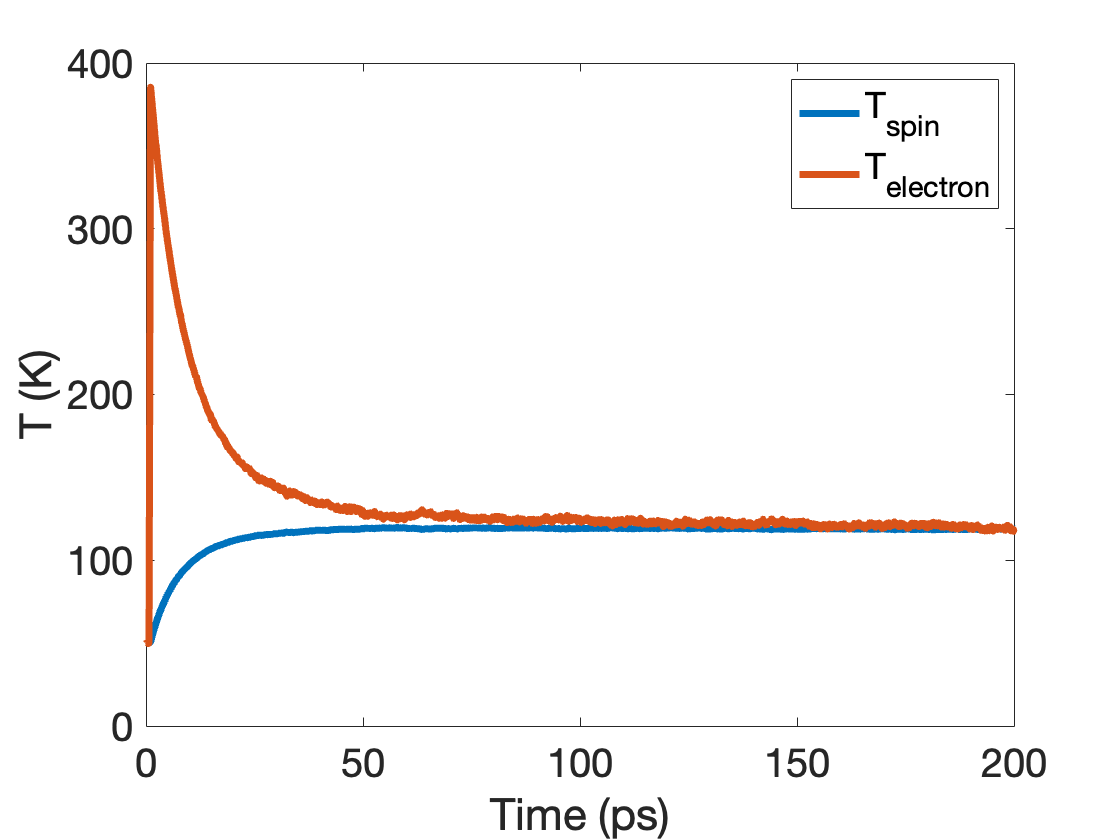}
\caption{Temperature dynamics of TbRh$_2$Si$_2$ for a laser fluence 375 J/m$^2$, $\alpha=0.001$. \label{fig:Tb_temp} }
\end{figure}

To get a deeper understanding the ultrafast demagnetization in \textit{Ln}Rh$_2$Si$_2$, we performed atomistic spin dynamics simulations starting from temperature 50 K. In the spin dynamics simulations we used the same value of the Gilbert damping for all materials. The value of the Gilbert damping was chosen to be $0.001$ \onlinecite{frietsch2015disparate} which is a realistic value for antiferromagnets and with this choice one obtains simulated magnetization curves that are close to experimentally observed values. We use the same Gilbert damping $\alpha$ value for all materials, because it was shown in Ref. \onlinecite{windsor2022exchange}
that Gilbert damping does not scale with de Gennes factor and it is expected that the valence band electronic structure is similar for all systems investigated here, as Fig.~\ref{fig:DOS_LnRh2Si2} also confirms. 
It should be noted that we have in fact studied the influence of Gilbert damping values in the range $0.001$--$0.05$, and we can confirm that in this range, the choice of Gilbert damping does not change the main conclusions presented in this work.
Typical spin and electron  temperatures from our simulations are presented in Fig.\ref{fig:Tb_temp} for TbRh$_2$Si$_2$. The lattice temperature remained fixed during our simulations (data not shown), as discussed above. The qualitative comparison of obtained spin and electronic temperatures with ones reported using HC3TM in \onlinecite{PhysRevB.106.174407, pankratova2024coupled} shows the similarities in behavior, such as rapid growth of electronic temperature following the absorption of the pulse, then slower growth of spin temperature.  Based on how  effective temperature models are constructed,  disregarding the lattice subsystem should have the similar effect as increasing the laser pulse or, alternatively, to a case of vanishing lattice damping. Additional tests (data not shown) on \textit{fcc} Ni confirm this behaviour. The other features of the magnetization curve remain the same. We show in subsection \ref{HC3TM} that HC2TM is able to reproduce the main features of system dynamics, even if the changes of the lattice temperature are disregarded.

\begin{figure}[h]
\includegraphics[width=0.45\textwidth]{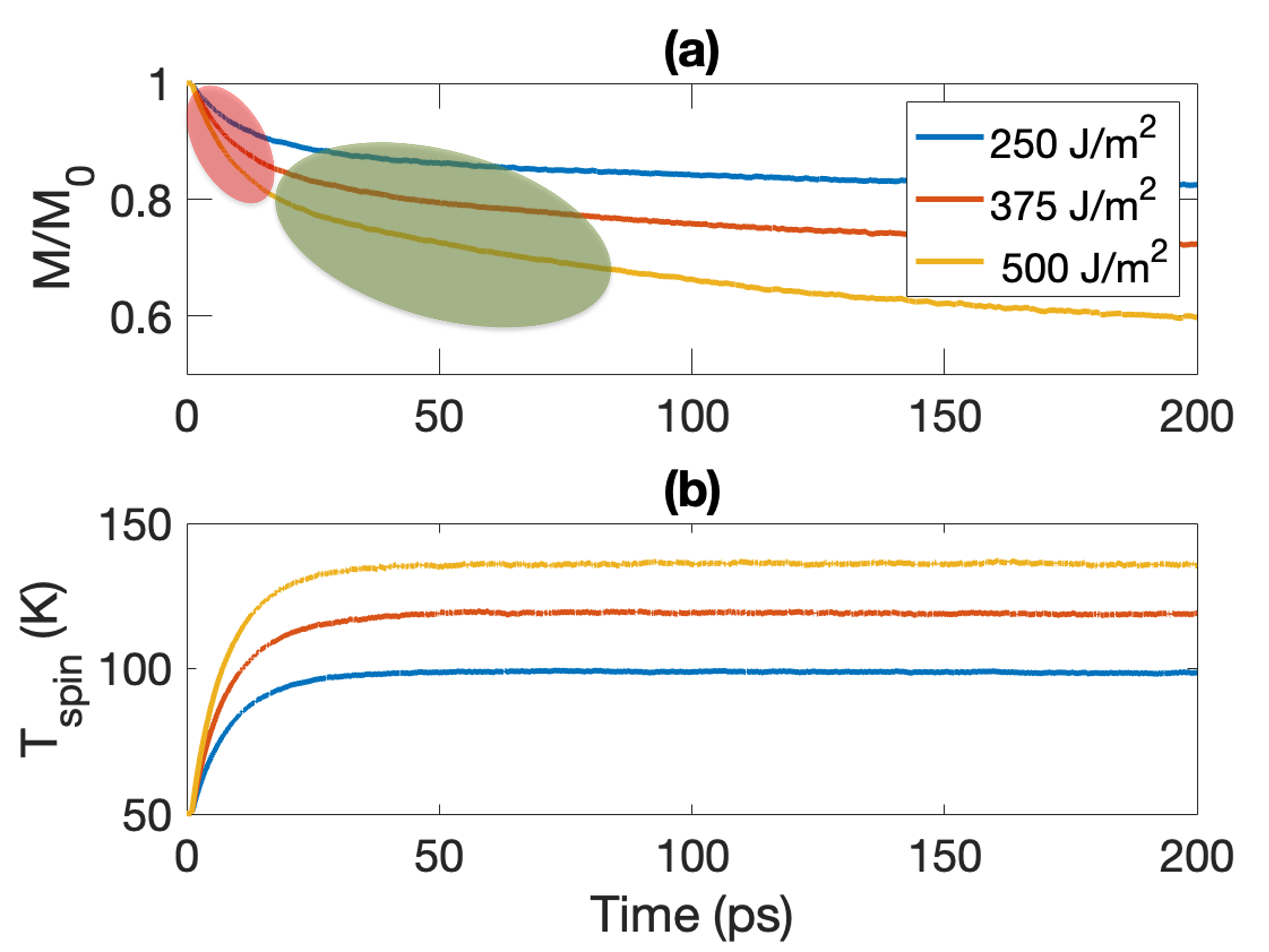}
\caption{Magnetization dynamics (a) and spin temperature dynamics (b) of TbRh$_2$Si$_2$ for various laser fluences. Shaded areas indicate  fast (red) and slow (green) parts of demagnetization process.\label{fig:Tb_fluence} }
\end{figure}

Calculated ultrafast magnetization dynamics curves of TbRh$_2$Si$_2$ are shown in Fig.\ref{fig:Tb_fluence}a for several values of laser fluence.
It can be seen from the Fig. \ref{fig:Tb_fluence}a that the demagnetization process consists of two main parts. It begins with a fast magnetization drop (marked by red shaded area), followed by a second slower decay (shown by green shaded area). In addition, as the figure shows, the first part of the demagnetization process becomes faster for higher laser fluence, something which is in agreement with experimental observations \onlinecite{windsor2022exchange}.  
The other studied materials exhibit similar behavior: first fast demagnetization, followed by a slower demagnetization. In addition, we present the dynamics of spin temperatures for various fluences shown in Fig.\ref{fig:Tb_fluence}b, which demonstrates the same trend as magnetization dynamics, specifically, the first sharp rise of the spin temperatures, that is followed by slower increase and thermalisation. 

\begin{figure}[h]
\begin{tabular}{cc}
(a) & \includegraphics[width=0.45\textwidth]{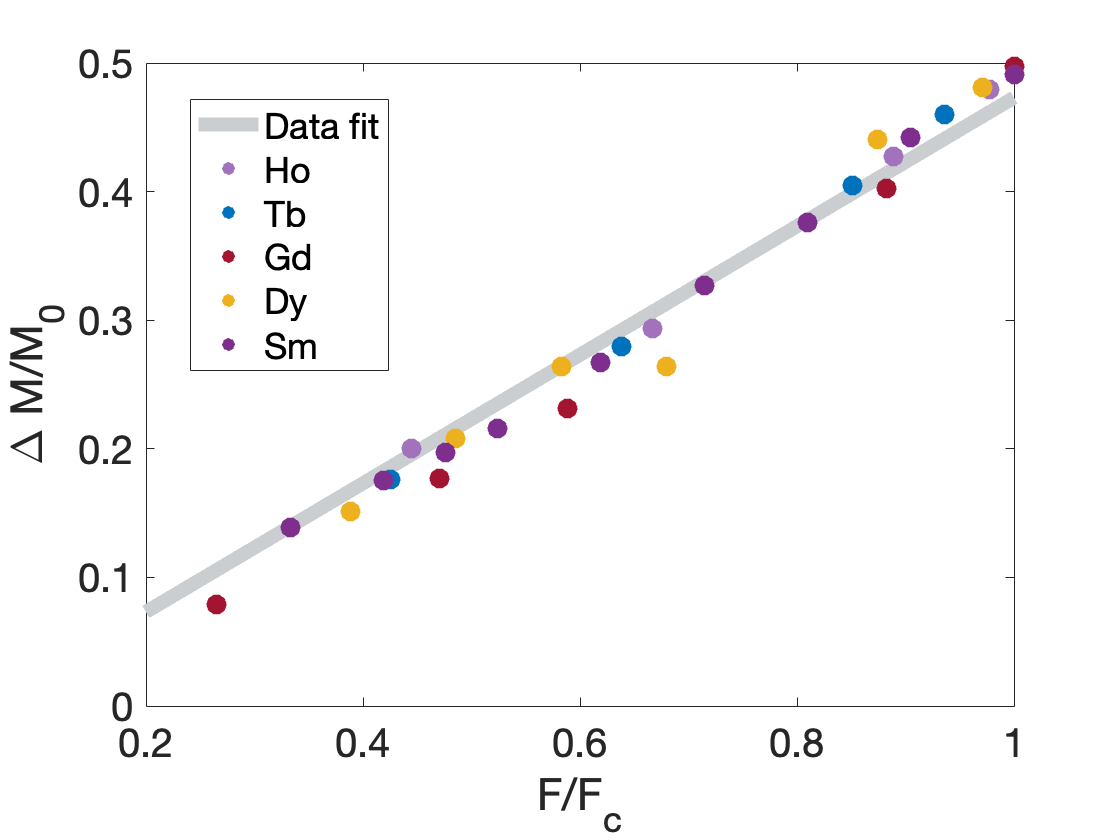}\\
(b) &\includegraphics[width=0.45\textwidth]{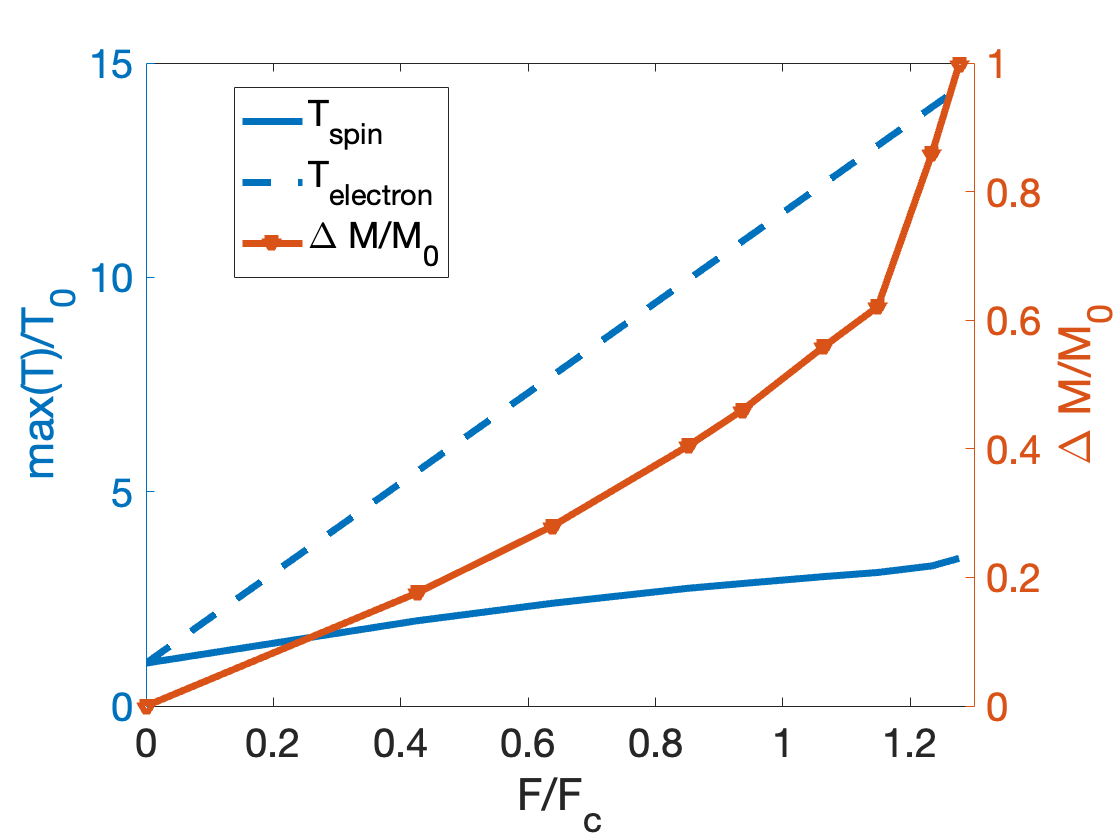}

\end{tabular}
\caption{(a) Total demagnetization
amplitude as a function of normalized laser fluence. Line is a guide for the eye. (b) Normalised spin and electron temperatures and total demagnetization amplitude as a function of a laser fluence for TbRh$_2$Si$_2$.\label{fig:fluence} }
\end{figure}

Next, we investigate the impact of laser fluence on the total demagnetization amplitude for all materials. Since the considered materials have a big difference in $T_N$, and, therefore, required different fluences for demagnetization, we adopt the concept of a critical laser fluence $F_c$ from Ref.~\onlinecite{windsor2022exchange}, which is the laser fluence leading to demagnetization amplitude $\Delta M/M_0=0.5$ (for example 600 J/m$^2$ is the $F_c$ for TbRh$_2$Si$_2$). The total demagnetization amplitude for all materials as a function of normalized fluence (i.e. laser fluence divided by the critical laser fluence) is presented in Fig.\ref{fig:fluence}a, and this is the main result of the present study. It can be seen from Fig.\ref{fig:fluence}a that the results of our simulations in general agree with experimental observations (see Fig.~2d in Ref.~\onlinecite{windsor2022exchange}) where a linear relationship between $\Delta M/M_0$ 
and normalized fluence was observed for this set of systems.

One can notice that for higher fluences or/and demagnetization amplitudes the experimental results start to deviate from linear dependence (see Fig.~2d of Ref.\onlinecite{windsor2022exchange}), especially for the Tb-system. We observe the same behavior in our simulations, especially for the Tb- and Dy-based systems (see Fig.~\ref{fig:fluence}a). We would like to note also, that in Ref.\onlinecite{lee2022robust}  for fluences substantially larger than $F_c$, this behavior goes in the other direction (saturation). 

To further analyze this non-linear behavior we performed simulations for higher fluences than those used in the experiments of Ref.\onlinecite{windsor2022exchange}. The results of these simulations are shown in Fig.~\ref{fig:fluence}b, where we focus on the electron- and spin temperature as well as $\Delta M/M_0$. It can be seen from Fig.~\ref{fig:fluence}b that while electronic and spin temperatures increase linearly with laser fluence for all fluences used in the simulations, the demagnetization amplitude demonstrates a linear behaviour only for smaller fluences $F$. In particular, for $F > F_c$ the deviation from a linear behaviour of $\Delta M/M_0$ is quite noticeable. Since the temperatures calculated in the simulations behave linearly with fluence, it is possible to relate the time dependent $\Delta M/M_0$ curve of Fig.~\ref{fig:fluence}b with a transient $\Delta M/M_0$ curve versus temperature (and not fluence) that has the same shape as the static curve shown in Fig.~\ref{fig:fluence}b. Hence, we find that the transient $\Delta M/M_0$ curve of Fig.~\ref{fig:fluence}b has strong similarities with the static $M(T)$ curve presented in Fig.~\ref{fig:M_temp} which suggests that the out of equilibrium magnetisation of the presently studies materials simply follows a static $M(T)$ curve at low fluences.
Over the picosecond times scales that are relevant for the dynamics of this class of materials, the simulations suggest that a quasi equilibrium is reached for the distribution of atomic moments and this quasi equilibrium evolves in time following the electronic temperature obtained in the simulations.

\subsection{Comparing the HC2TM and HC3TM for GdRh$_2$Si$_2$}
\label{HC3TM}

To investigate the importance of lattice dynamics for the ultrafast demagnetisation, we perform coupled atomistic spin-lattice dynamics simulations of GdRh$_2$Si$_2$, and apply the heat-conserving three-temperature model (HC3TM), which was proposed and described in detail in Refs.\onlinecite{pankratova2024coupled, PhysRevB.106.174407}. In these simulations, we do not include spin-lattice coupling specifically in the Hamiltonian. Instead coupling between spin- and lattice subsystems takes place as flow of heat between the two systems, in part mediated by the electron temperature (for details see Refs.\onlinecite{pankratova2024coupled, PhysRevB.106.174407}). It was shown in Ref.\onlinecite{pankratova2024coupled} that even in the absence of explicit spin-lattice coupling in the Hamiltonian (e.g. as introduced in Ref.\onlinecite{PhysRevB.99.104302}), lattice dynamics can play an important role in ultrafast demagnetisation, impacting demagnetisation amplitude and remagnetisation. Details of lattice dynamics simulations, needed for the HC3TM, are presented in Appendix \ref{append:Force}. This includes the equations of motion, lattice damping and calculations of the force constant. In these simulations, the spin-dynamics is treated according to Eqns. \eqref{e:Heisenberg_model} and \eqref{eq1}. 

\begin{figure}[h]

\includegraphics[width=0.48\textwidth]{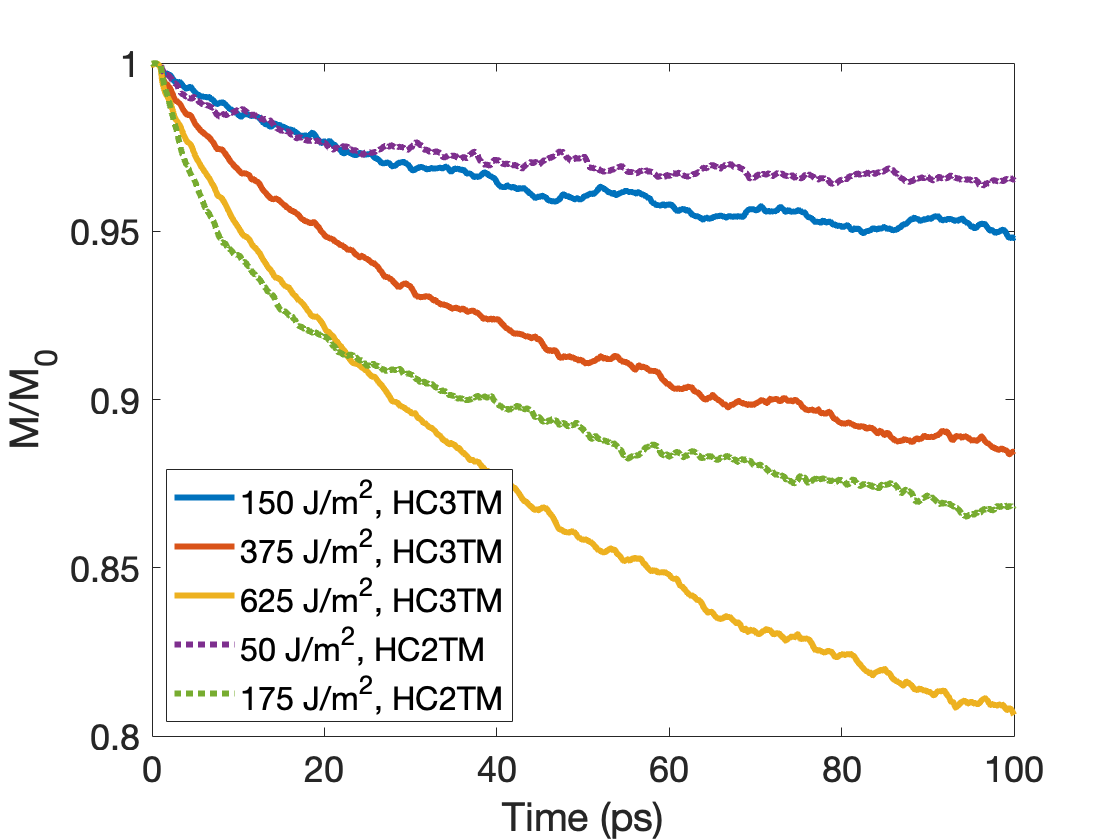}\\
\caption{Magnetisation dynamics of GdRh$_2$Si$_2$ in HC3TM (solid lines) and HC2TM (dash lines) for various laser fluences. 
\label{fig:lattice_impact_fluence} }
\end{figure}

We start by studying ultrafast magnetisation dynamics for various fluences, and compare results of HC3TM with HC2TM for GdRh$_2$Si$_2$. It can be seen from Fig.\ref{fig:lattice_impact_fluence} that account of lattice dynamics, as provided in the HC3TM, leads to slower magnetisation dynamics especially during the first 10 to 20 ps. Moreover, while comparing HC2TM and HC3TM for the same (or similar) fluence, we observe that HC2TM leads to higher demagnetisation amplitude.  
Interestingly, very similar demagnetisation rate during the first 10 to 20 ps are observed for the HC2TM and HC3TM, provided one considers different fluences. For instance, the 175 J/m$^2$ HC2TM data are almost on top of the 625 J/m$^2$ HC3TM data, and the 50 J/m$^2$ HC2TM data lie almost on top of the 150 J/m$^2$ HC3TM data. Hence, for the present system it seems that a scaling of fluence with a factor 3-4 provides very similar results for simulations of HC2TM and HC3TM. In general the magnetisation curves in both models leads to similar features of the magnetisation dynamics, such as faster magnetisation drop and then slower demagnetisation on longer timescales, which justify the use of HC2TM and the results provided in Fig.7. The typical example of temperature dynamics for HC3TM can be found in Appendix \ref{App:temp}. 

\begin{figure}[h]
\includegraphics[width=0.47\textwidth]{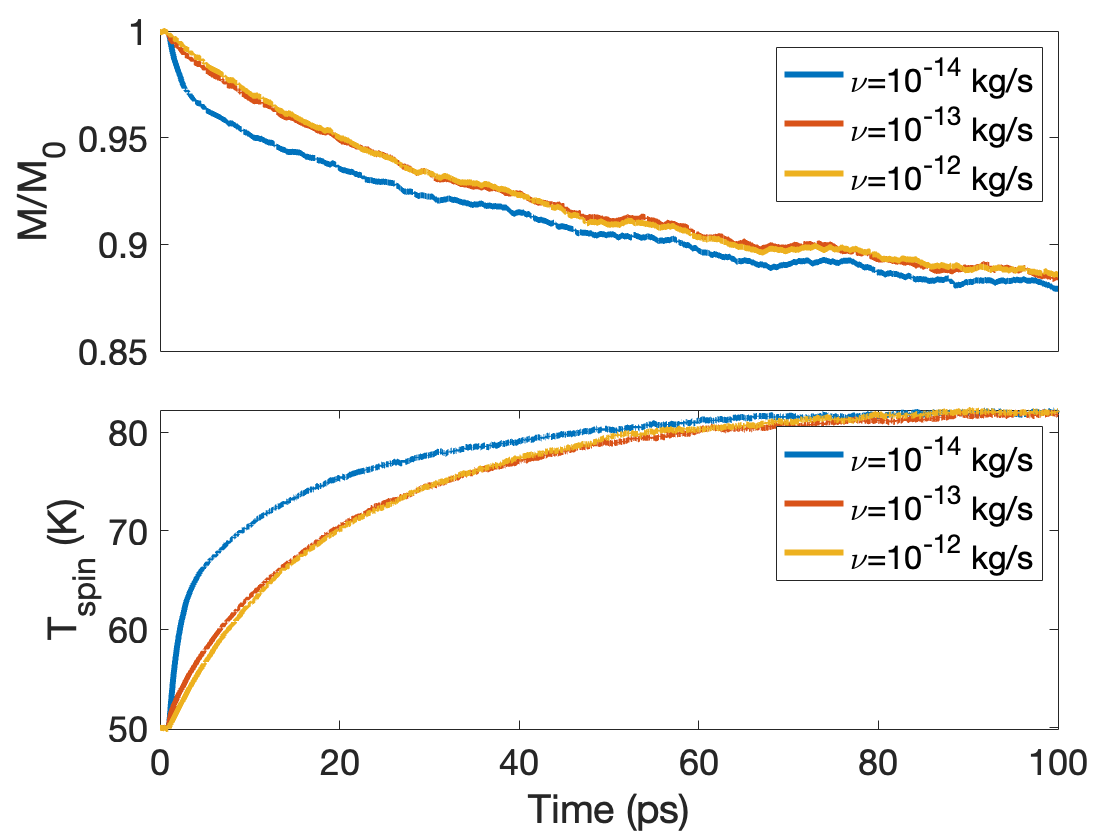}
\caption{ (top panel) Magnetisation dynamics of GdRh$_2$Si$_2$ in HC3TM (solid lines) for various values of lattice damping (laser fluence 375 J/m$^2$) in comparison with magnetisation dynamics of GdRh$_2$Si$_2$ in HC2TM. (bottom panel) Corresponding spin temperature dynamics for various values of lattice damping (laser fluence 375 J/m$^2$).
\label{fig:lattice_impact_damping} }
\end{figure}

The transfer of heat between electrons and lattice is tuned by several interactions, one being the lattice damping $\nu$ parameter (see Appendix for details and Refs. \onlinecite{pankratova2024coupled,PhysRevB.106.174407}). It can be seen in Fig.\ref{fig:lattice_impact_damping}(b) that smaller lattice damping leads to faster demagnetisation and slightly bigger demagnetisation amplitude, which is consistent with previous results for 3d ferromagnets\cite{pankratova2024coupled}. This trend also explains faster demagnetisation and bigger demagnetisation amplitudes in HC2TM, compared to HC3TM, which can be considered as a limit of zero lattice damping. Increase of demagnetisation amplitude with reduction in lattice damping is connected to a higher amount of heat transferred from electrons to the spin system and, therefore, the higher spin temperatures, as can be see at Fig.\ref{fig:lattice_impact_damping}b.

\begin{figure}[h]
\includegraphics[width=0.47\textwidth]{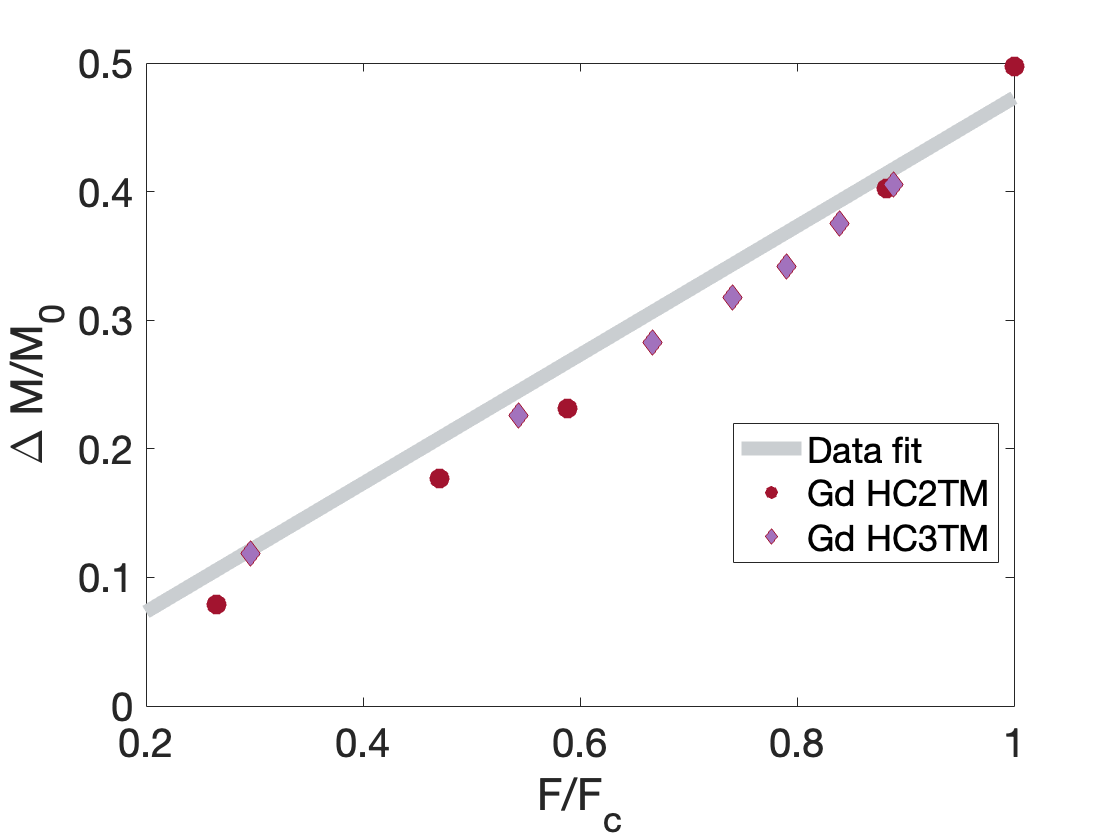}
\caption{Total demagnetization amplitude in HC2TM (red circles) and HC3TM (violet diamonds) for GdRh$_2$Si$_2$. 
\label{fig:lattice_impact_linear} }
\end{figure}

We complete our analysis of how the lattice dynamics impact the magnetization dynamics by considering the demagnetization amplitude as a function of normalized laser fluence in the HC3TM and comparing it with values obtained in HC2TM. This is presented in Fig.\ref{fig:lattice_impact_linear}. We would like to note that F$_c$ for GdRh$_2$Si$_2$ is different in HC3TM and HC2TM, due to the differences in the simulations, as discussed above. Taking this difference into account, it can be seen from Fig.\ref{fig:lattice_impact_linear} that HC2TM and HC3TM lead to a very similar linear relationship between magnetization amplitude and normalised fluence. As can be seen, details of the lattice dynamics do not impact this linear trend at low fluences, confirming the suitability of HC2TM for studying ultrafast demagnetisation dynamics in the present system and for validation of the results shown in Fig.7.

\section{Conclusions}

In this work we give a thorough account of the theoretical results of the electronic and magnetic properties of the \textit{Ln}Rh$_2$Si$_2$ system, based on DFT coupled to a method for evaluating spin Hamiltonian parameters. \cite{LKAG1987,Szilva2023} From the spin Hamiltonian, an effective Weiss field can be calculated that is here used in atomistic spin-dynamics simulations \onlinecite{Eriksson2017} coupled to a heat conserving temperature model \cite{PhysRevB.106.174407} that enables calculations of transient temperature profiles relevant for simulations of pump-probe experiments. 

The simulations presented here for \textit{Ln}Rh$_2$Si$_2$ agree with experimental observations of pump-probe experiments, especially for lower fluences, where in Ref.\onlinecite{windsor2022exchange} the demagnetization amplitude as a function of fluence was studied for amplitudes smaller than $F/F_c=0.5$. The linear dependence of demagnetization amplitude on laser fluence is widely reported for various materials both from theoretical and experimental studies \cite{lee2024controlling,mishra2021ultrafast,PhysRevB.81.174401,windsor2022exchange,pankratova2024coupled}, however, the motivation behind this linear trend has hitherto been only little discussed and therefore it remains poorly understood. In Ref.\onlinecite{lee2024controlling} it was recently reported that linear scaling of demagnetisation amplitude with normalized fluence is related to the linear scaling of N\'eel temperature with normalized fluence for lanthanide-based antiferromgants. It was reported in Ref.~\onlinecite{pankratova2024coupled} (see Fig.~4 in  Supplementary Information of that paper) that for elemental ferromagnets the spin temperature and transient magnetization during the ultrafast magnetization dynamics deviate very little from the equilibrium $M(T)$ curve. Our findings here for antiferromagnetic compounds are consistent with the observations of Ref.~\onlinecite{pankratova2024coupled}. Hence a picture that emanates from these works, as well as of Ref.~\onlinecite{PhysRevResearch.2.013180}, 
is that the transient magnetization can be described as resulting from a quasi-equilibrium distribution of the atomic moments, and that this distribution evolves in time as a result of the changes of the temperature. It should be noted here that it is primarily for these simulations that the electronic temperature is relevant, since in heat conserving temperature models for transient magnetism the electronic temperature enters the fluctuating field that changes the magnetic state.   

Finally we note that from the simulations put forth here, one predicts that for higher fluences the linearity between the demagnetization amplitude and fluence becomes broken; for values of $F/F_c > 1.0$ a non-linear behaviour is quite distinct. It was shown that this result does not depend on accounting or disregarding lattice dynamics in simulations. These predictions can hopefully be confirmed or refuted in experimental work, and would be welcome since they could give creedence to the theoretical model proposed here. 

\section{Acknowledgements}
This work was financially supported by the Knut and Alice Wallenberg Foundation through grant numbers 2018.0060, 2021.0246, and 2022.0108. OE and MP acknowledge support from the Wallenberg Initiative Materials Science for Sustainability (WISE) funded by the Knut and Alice Wallenberg Foundation (KAW). 
OE also acknowledges financial support from the Swedish Research Council (Vetenskapsrådet, VR), the European Research Council (854843-FASTCORR), and STandUP. MP acknowledge support from Olle Engkvist foundation and from WISE-WASP joint call. OE and AB acknowledge eSSENCE for financial support.
DT acknowledges financial support from the Swedish Research Council (Vetenskapsrådet, VR) grant numbers 2019-03666 and 2023-04239. AE acknowledges the funding by the Fonds zur F\"orderung der Wissenschaftlichen Forschung (FWF) under Grant No. I 5384. 
LR and YWW acknowledge funding from the Deutsche Forschungsgemeinschaft (DFG, German Research Foundation) within the Transregio TRR 227 - 328545488 Ultrafast Spin Dynamics (Projects No. A10 and B07).
This publication is part of the project NL-ECO: Netherlands Initiative for Energy-Efficient Computing (with project number NWA. 1389.20.140) of the NWA research programme Research Along Routes by Consortia which is financed by the Dutch Research Council (NWO). 
The computations were enabled by resources provided by the National Academic Infrastructure for Supercomputing in Sweden (NAISS) at the National Supercomputing Centre (NSC, Tetralith cluster) partially funded by the Swedish Research Council through grant agreement no.\,2022-06725.
Structural sketch in Fig.~\ref{fig:DOS_LnRh2Si2} was produced using the \textsc{VESTA3} software \cite{vesta}.

\bibliographystyle{apsrev4-2}
\bibliography{main}

\clearpage
\appendix

\section{Force constants calculation}
\label{append:Force}
We used the Vienna Ab-initio Simulation Package (VASP) \onlinecite{kresse1996efficient} and PHONOPY \onlinecite{togo2020phonon} to compute the force constants required for spin-lattice dynamics simulations. First, we used a two-formula-unit tetragonal unit cell to perform DFT calculations using the projected augmented wave (PAW) method \onlinecite{kresse1999ultrasoft}, as implemented in the Vienna Ab-initio Simulation Package (VASP) \onlinecite{kresse1996efficient}. The exchange-correlation functional was described using the generalized gradient approximation (GGA) with the Perdew-Burke-Ernzerhof (PBE) parametrization \onlinecite{perdew1996generalized}. A plane-wave energy cutoff of 500 eV was applied for the basis set, along with a $\Gamma$-centered Monkhorst-Pack k-mesh of 20×20×10 to ensure convergence of the total energy and local magnetic moments. To accurately capture the localized nature of Gd $4f$ electrons, we employed DFT+$U$ calculations using the Hartree-Fock approximation \onlinecite{kotliar2006electronic}. The rotationally invariant formulation by Liechtenstein et al. \onlinecite{liechtenstein1995density} was applied with Coulomb interaction parameters $U$=6.7 eV, and $J$=0.7 eV for the Gd $4f$ states, based on prior studies \onlinecite{guttler2016robust}. Before calculating the force constants, the experimental unit cell \onlinecite{windsor2022exchange} was fully relaxed, optimizing atomic positions, cell shape, and volume to meet a force convergence criterion of 1~meV/A. Force constants for the relaxed structure were computed using PHONOPY \onlinecite{togo2020phonon} and density functional perturbation theory (DFPT), as implemented in VASP \onlinecite{kresse1996efficient}. The force constant matrix elements (\( \text{eV/Å}^2 \)) are tabulated in Table~\ref{tab:force_constants}, with Gd at position (0, 0, 0) in the two-formula-unit tetragonal cell containing 10 atoms taken as the reference. The atomic positions of Gd, Rh, and Si in the two-formula-unit tetragonal cell are as follows. The cell contains 2 Gd atoms, 4 Rh atoms, and 4 Si atoms. The positions are given in direct coordinates: \\
\begin{center}
1. \textbf{Gd:} (0.00000, 0.00000, 0.00000) \\ 
2. \textbf{Gd:} (0.50000, 0.50000, 0.50000) \\ 
3. \textbf{Rh:} (0.00000, 0.50000, 0.25000) \\ 
4. \textbf{Rh:} (0.00000, 0.50000, 0.75000) \\ 
5. \textbf{Rh:} (0.50000, 0.00000, 0.25000) \\ 
6. \textbf{Rh:} (0.50000, 0.00000, 0.75000) \\ 
7. \textbf{Si:} (0.00000, 0.00000, 0.37742) \\ 
8. \textbf{Si:} (0.00000, 0.00000, 0.62258) \\ 
9. \textbf{Si:} (0.50000, 0.50000, 0.87742) \\ 
10. \textbf{Si:} (0.50000, 0.50000, 0.12258)
\end{center}

\begin{figure}[htb]
    \includegraphics[width=0.45\textwidth]{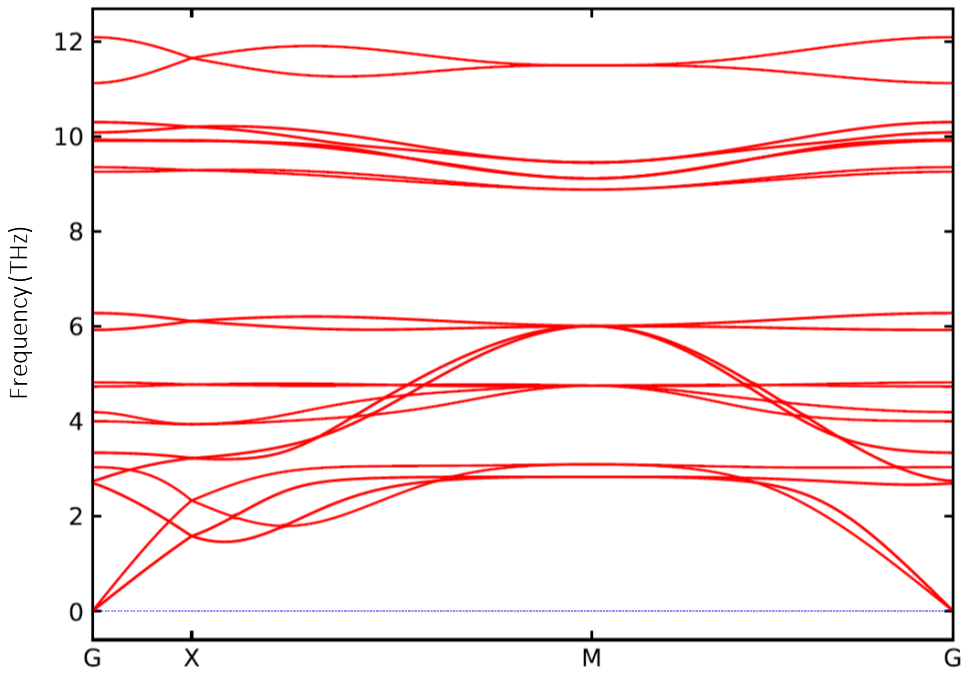}
    \caption{Phonon band structure of GdRh$_2$Si$_2$.}
    \label{fig:Phonon_GdRh2Si2}
\end{figure}

The phonon band structure, shown in Fig. \ref{fig:Phonon_GdRh2Si2}, confirms the dynamical stability of the system, with no imaginary phonon frequencies observed.

\begin{table}[h!]
    \centering
    \caption{Force constant matrix elements for each pair \(i\)-\(j\) in the two-formula-unit cell.}
    \label{tab:force_constants}
    \begin{tabular}{cccc}
        \toprule
        Pair \(i\)-\(j\) & \(\Phi_{xx}\) & \(\Phi_{xy}\) & \(\Phi_{xz}\) \\ 
        & \(\Phi_{yx}\) & \(\Phi_{yy}\) & \(\Phi_{yz}\) \\ 
        & \(\Phi_{zx}\) & \(\Phi_{zy}\) & \(\Phi_{zz}\) \\ 
        \midrule
        1-1 & \(5.17472\) & \(0.00000\) & \(0.00000\) \\ 
            & \(0.00000\) & \(5.17472\) & \(0.00000\) \\ 
            & \(0.00000\) & \(0.00000\) & \(6.26490\) \\ 
        \midrule
        1-2 & \(0.01013\) & \(0.00000\) & \(0.00000\) \\ 
            & \(0.00000\) & \(0.01013\) & \(0.00000\) \\ 
            & \(0.00000\) & \(0.00000\) & \(0.28652\) \\ 
        \midrule
        1-3 & \(0.21722\) & \(0.00000\) & \(0.00000\) \\ 
            & \(0.00000\) & \(-0.59987\) & \(0.00000\) \\ 
            & \(0.00000\) & \(0.00000\) & \(-0.97715\) \\ 
        \midrule
        1-4 & \(0.21722\) & \(0.00000\) & \(0.00000\) \\ 
            & \(0.00000\) & \(-0.59987\) & \(0.00000\) \\ 
            & \(0.00000\) & \(0.00000\) & \(-0.97715\) \\ 
        \midrule
        1-5 & \(-0.59987\) & \(0.00000\) & \(0.00000\) \\ 
            & \(0.00000\) & \(0.21722\) & \(0.00000\) \\ 
            & \(0.00000\) & \(0.00000\) & \(-0.97715\) \\ 
        \midrule
        1-6 & \(-0.59987\) & \(0.00000\) & \(0.00000\) \\ 
            & \(0.00000\) & \(0.21722\) & \(0.00000\) \\ 
            & \(0.00000\) & \(0.00000\) & \(-0.97715\) \\ 
        \midrule
        1-7 & \(-0.35841\) & \(0.00000\) & \(0.00000\) \\ 
            & \(0.00000\) & \(-0.35841\) & \(0.00000\) \\ 
            & \(0.00000\) & \(0.00000\) & \(-0.91007\) \\ 
        \midrule
        1-8 & \(-0.35841\) & \(0.00000\) & \(0.00000\) \\ 
            & \(0.00000\) & \(-0.35841\) & \(0.00000\) \\ 
            & \(0.00000\) & \(0.00000\) & \(-0.91007\) \\ 
        \midrule
        1-9 & \(-1.84982\) & \(0.00000\) & \(0.00000\) \\ 
            & \(0.00000\) & \(-1.84982\) & \(0.00000\) \\ 
            & \(0.00000\) & \(0.00000\) & \(-0.41154\) \\ 
        \midrule
        1-10 & \(-1.84982\) & \(0.00000\) & \(0.00000\) \\ 
             & \(0.00000\) & \(-1.84982\) & \(0.00000\) \\ 
             & \(0.00000\) & \(0.00000\) & \(-0.41154\) \\ 
        \bottomrule
    \end{tabular}
\end{table}

\section{Atomistic spin-lattice dynamics simulations}

Coupled atomistic spin-lattice dynamics simulations are performed, using Langevin dynamics, similarly to Ref.\, \onlinecite{PhysRevB.99.104302,pankratova2024coupled,PhysRevB.106.174407}. In addition to spin dynamics governed by Eq.\ref{eq1}, the lattice dynamics is described by:

\begin{equation}
\label{eq2}
    \frac{d \boldsymbol{u}_k}{dt} = \boldsymbol{v}_k
\end{equation}

\begin{equation}
\label{eq3}
    \frac{d \boldsymbol{v}_k}{dt}= \frac{\boldsymbol{F}_k}{M_k}+\frac{\boldsymbol{F}_k^{fl}}{M_k} - \nu \boldsymbol{v}_k,
\end{equation}
where the lattice damping constant is denoted $\nu$, atomic displacements and velocities are  given by $\boldsymbol{u_k}$,$\boldsymbol{v_k}$ respectively. The force at site k is defined by $\boldsymbol{F}_k=-\partial H_{SLD}/\partial \boldsymbol{u}_k$. Spin-lattice Hamiltonian $H_{SLD}$ is a sum of magnetic contribution  given in Eq.\ref{e:Heisenberg_model} and lattice contribution $H_{SLD} =H+H_{LL}$, where $H_{LL}$ is:

\begin{equation}
    H_{\mathrm{LL}} = \frac{1}{2} \sum_{kl} \Phi_{kl}^{\mu \nu} u_k^{\mu} u_l^{\nu} + \frac{1}{2} \sum_{k} M_{k} \nu_k^{\mu} \nu_k^{\mu},
\end{equation}
where  $\Phi_{kl}^{\mu \nu}$ is the force constant tensor, and $M_k$ is the mass of atom $k$.

In these types of Langevin simulations one employs stochastic fields $\boldsymbol{F}_{k}^{fl}$, as white noise with properties $\langle F_{i,\mu}^{fl}(t) F_{j,\nu}^{fl}(t') \rangle=2D_L \delta_{kl}\delta_{\mu\nu}\delta(t-t')$. In our simulations, we use $D_L= \nu M k_B T$,  (please see e.g. Ref.\,\onlinecite{PhysRevB.99.104302,pankratova2024coupled,PhysRevB.106.174407}).

The lattice temperature  $T_l$ for the HC3TM is calculated from the average kinetic energy of the lattice vibrations; $\langle E^{kin}_l\rangle/k_B$.

\section{Spin, electron, and lattice temperatures in HC3TM}

\label{App:temp}

\begin{figure}[htb]
\includegraphics[scale=0.23]{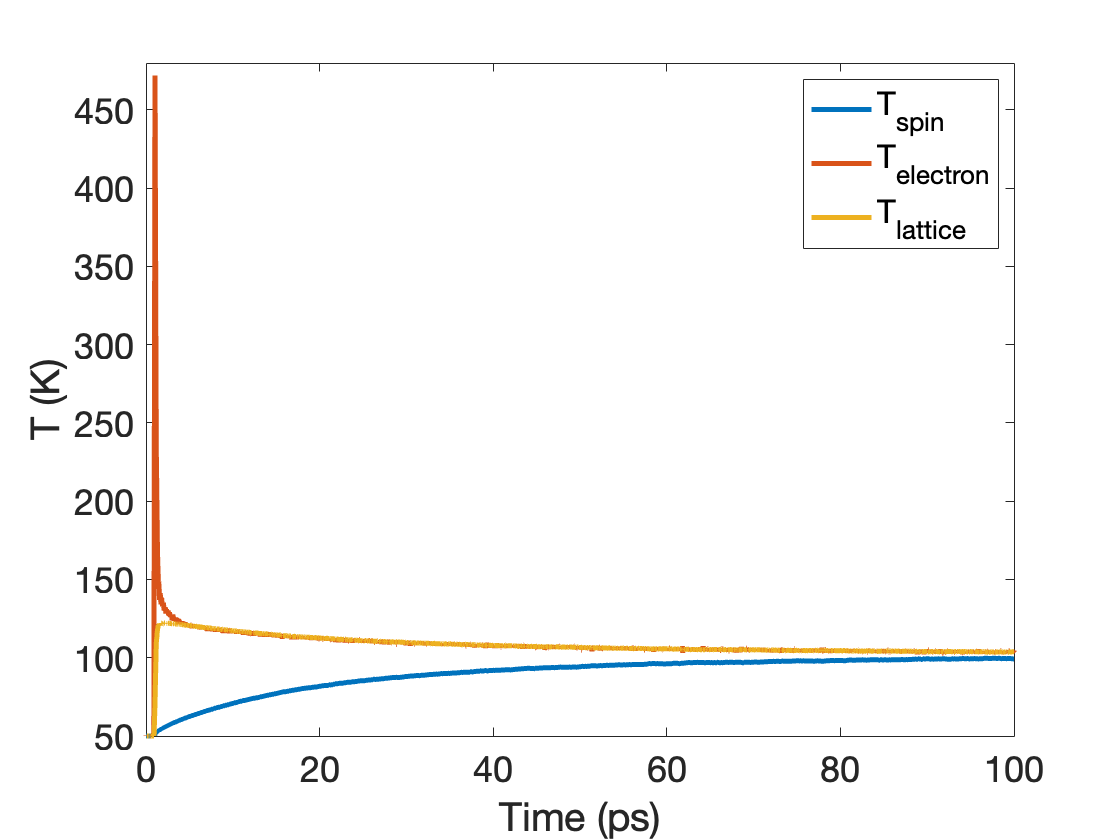}
\caption{Spin, lattice, and electron temperatures in HC3TM for GdRh$_2$Si$_2$}
\label{fig:HC3TM_temp}
\end{figure}

The dynamics of spin, lattice and electron temperatures in HC3TM for GdRh$_2$Si$_2$ is presented in Fig.\ref{fig:HC3TM_temp}. One can notice that the electron temperature is raising first, followed closely by lattice temperature, which is in line with previous experimental studies \cite{lee2022robust}, where the thermalisation of electronic and lattice temperatures was observed during the first picoseconds after the laser pulse. The rise of $T_e$ and $T_l$ is then followed by a much slower rise of a spin temperature. The slow rise of spin temperature, in comparison for example, with 3d ferromagnets \cite{pankratova2024coupled, PhysRevB.106.174407} is explained by a very low Gilbert damping, used in this work, which in the HC3TM model is responsible for a transfer of heat between electronic and spin subsystems \onlinecite{PhysRevB.106.174407,pankratova2024coupled}, as well as influenced by exchange interactions. Therefore, the difference in temperatures dynamics, in comparison with $3d$ ferromagnets \onlinecite{pankratova2024coupled, PhysRevB.106.174407} previously studied in HC3TM, is explained by different Gilbert damping values, different force constants and exchange interactions, which influence magnetization dynamics and, therefore, the temperatures, "measured" during the atomistic spin-lattice dynamics simulations. 

\end{document}